\documentclass[twocolumn,reprint]{revtex4-1}
\usepackage{graphicx}
\usepackage{times}
\usepackage{bm}
\usepackage{amsmath,amssymb}

\begin{document}
\title{Vertex-type thermal correction to the one-photon transition rates}

\author{D. Solovyev}
\email[E-mail:]{d.solovyev@spbu.ru}
\affiliation{ Department of Physics, St. Petersburg State University, Petrodvorets, Oulianovskaya 1, 198504, St. Petersburg, Russia }
\author{T. Zalialiutdinov}
\affiliation{ Department of Physics, St. Petersburg State University, Petrodvorets, Oulianovskaya 1, 198504, St. Petersburg, Russia }
\author{A. Anikin}
\affiliation{ Department of Physics, St. Petersburg State University, Petrodvorets, Oulianovskaya 1, 198504, St. Petersburg, Russia }

\begin{abstract}
Thermal corrections to the one-photon spontaneous and induced transition probabilities for hydrogen and hydrogen-like ions are evaluated. The found thermal corrections are given by the vertex Feynman graph, where the vertex represents the thermal interaction between the bound electron and the nucleus. All derivations of thermal corrections to bound-bound transitions for an atom exposed to blackbody radiation (BBR) are made in a fully relativistic approach within the framework of the adiabatic $S$-matrix formalism. It is found that the vertex-type radiative corrections to the transition rates can be at the level of a few percent to corresponding spontaneous rates for highly excited states in the hydrogen atom. A comprehensive analysis of the vertex-type thermal corrections for hydrogen-like atomic systems is presented.
\end{abstract}

\maketitle

\section{Introduction}

Since the early days of quantum mechanics (QM), the study of absorption and emission of photons by atomic systems played a key role in the development of modern quantum field theory and its practical application in various fields of physics, chemistry, engineering and etc. Further development of quantum mechanics led to the creation of a quantum electrodynamical description (QED) of light interaction with matter, within which all processes are described in terms of scattering cross-sections and transition rates. Subsequent experimental observations and their growing accuracy have required taking into accounting more complex effects, in particular, radiative QED corrections. In this regard, a detailed theoretical analysis of various radiative corrections providing a versatile verification of fundamental physics is needed. The precise values of the transition rates in various atomic systems are also of interest for studying the processes of an atomic collision or interpreting spectra from astrophysical sources \cite{Dubrovich_2020,RubChSun}. Moreover, accurate calculations of the transition rates can serve for the verification of basic parts in more complicated processes, such as, e.g., the parity violation amplitudes in heavy ions and atoms \cite{PhysRevA.67.052110,PhysRevA.72.062105}. 


To study the above examples, it becomes extremely important to determine accurately the lifetimes and decay rates of different atomic systems. Since real physical processes always occur in the presence of certain external fields, it is necessary to study theoretically their influence on the spectral characteristics of atoms. One of these fields is represented by a blackbody radiation (BBR). In astrophysics, an external field with a mostly Planck spectrum (cosmic microwave background, radiation from powerful sources, nebulae, etc.) affects the population of atomic states, which leads to significant changes in the dynamic of astrophysical processes. Under laboratory conditions, the radiation of blackbody radiation also has impact on the physics of the studied processes and, consequently, requires detailed study \cite{Farley,Hall,Saf1,SKC,Deg}. A rigorous analysis of these effects is possible within the framework of the quantum electrodynamics theory at finite temperatures (TQED) for bound states, see, for example, \cite{S-2020} and references therein.

Concentrating on the electric and magnetic dipole transitions in the H-like ions, we extend the approach developed in \cite{S-2020,ZAS-2l,ZSL-1ph} to the study of the lowest-order thermal radiative corrections to decay rates given by Feynman diagram Fig.~\ref{fig1}. Below we demonstrate that these corrections can be significant for the measurements of lifetimes of highly excited (Rydberg) states in these atomic systems since their relative magnitude can reach a level of several percent at room temperature. The interaction potential, Fig.~\ref{fig2}, which shifts atomic energy levels and corrects wave function of the bound electron, was recently derived in \cite{S-2020}. It was found that thermal potential, Fig.~\ref{fig2}, produces the dominant thermal frequency shift arising in a heat bath and could exceed well-known BBR-induced Stark shift \cite{SZA-2020}.

The paper is organized as follows. In section~\ref{sectionA} we start from a brief description of the Gell-Mann and Low adiabatic formalism used to evaluate the one-photon transition rates and radiative corrections. The derivation of master equations for vertex-type correction to the transition rate is given in section~\ref{sectionB}. In section~\ref{sectionB} the nonrelativistic limit of derived equations is also presented. All the derivations are performed within the framework of rigorous quantum electrodynamics theory at finite temperatures and are applicable for the H-like ions. The results of numerical calculations of the thermal vertex-type corrections for electric and magnetic dipole transitions in H-like ions are discussed in section~\ref{theend}. The relativistic units $ \hbar=m_{e}=c=1 $ ($ m_{e} $ is the electron rest mass, $c$ is the speed of light and $\hbar$ is the reduced Planck constant) are used throughout the paper.

\section{Adiabadic $S$-matrix formalism}
\label{sectionA}

For the description of radiative QED correction we start from the basic formulas of the Gell-Mann and Low adiabatic formalism \cite{Gell}. Within this approach the initial formula for the energy shift $\Delta E_A$ of an excited atomic state $A$ is 
\begin{eqnarray}
\label{1}
\Delta E_A = \lim_{\eta\rightarrow 0}\frac{1}{2}i\eta\frac{e\frac{\partial}{\partial e}\langle A|\hat{S}_{\eta}|A\rangle }{\langle A|\hat{S}_{\eta}|A\rangle }.
\end{eqnarray}
The adiabatic $S$-matrix $\hat{S}_{\eta}$ in Eq. (\ref{1}) differs from the orinary $S$-matrix by the presence of the exponential factor $e^{-\eta|t|}$ in each (interaction) vertex. It refers to the concept of adiabatic switching on and off the interaction introduced formally by the replacement $\hat{H}_{{\rm int}}(t) \longrightarrow \hat{H}^\eta_{{\rm int}}(t) = e^{-\eta|t|}\,\hat{H}_{{\rm int}}(t)$. The symmetric version of the adiabatic formula containing $S_{\eta}(\infty,-\infty)$, which is more convenient for the QED calculations, was proposed by Sucher \cite{Sucher}. The first application of the formula (\ref{1}) to calculations within bound-state QED was made in \cite{Lab}. In \cite{Lab} it was shown how to deal with the adiabatic exponential factor when evaluating the real part of corrections to the energy levels Eq. (\ref{1}) (see also \cite{LabKlim}). In this section, we employ the same methods for evaluating the imaginary part of Eq. (\ref{1}), see \cite{LSP-sep}.

For a free atom (or ion) in the state $|A\rangle$ interacting with the photon vacuum $|0_\gamma\rangle$ (i.e. $|A, 0_\gamma\rangle =|A\rangle|0_\gamma\rangle$ in the absence of external fields) the complex energy correction contains only diagonal $S$-matrix elements of even order, since $\langle 0_\gamma|\hat{S}^{(1)}_{\eta}|0_\gamma\rangle = \langle 0_\gamma|\hat{S}^{(3)}_{\eta}|0_\gamma\rangle = 0$ etc. For the separation of the imaginary part of the energy shift $\Delta E_A^{(2i)}$ of a given order $2i$, it is more convenient to represent Eq. (\ref{1}) in terms of a perturbation series of the form (up to terms $e^4$) \cite{LabKlim}
\begin{eqnarray}
\label{2}
\Delta E_A = \lim_{\eta\rightarrow 0} i\eta\, \left[\langle A|\hat{S}^{(2)}_{\eta}|A\rangle + 
\right.
\\
\left.
\nonumber
\left(2 \langle A|\hat{S}^{(4)}_{\eta}|A\rangle - \langle A|\hat{S}^{(2)}_{\eta}|A\rangle^2\right)
+\dots \right].
\end{eqnarray}

For the adiabatic $\hat{S}_{\eta}$ matrix the standard expansion in powers of the interaction constant $e$ was used
\begin{eqnarray}
\label{3}
\hat{S}_{\eta}(\infty,-\infty)=1+\sum\limits_{i=1}^{\infty}\hat{S}^{(i)}_{\eta}(\infty,-\infty).
\end{eqnarray}
To separate real and imaginary parts of the matrix elements at any given order of perturbation theory, one can write
\begin{eqnarray}
\label{4}
\langle A|\hat{S}^{(i)}_{\eta}|A\rangle  = \mathrm{Re}\langle A|\hat{S}^{(i)}_{\eta}|A\rangle +i\mathrm{Im}\langle A|\hat{S}^{(i)}_{\eta}|A\rangle.
\end{eqnarray}
The only one second-order term describes the pure one-photon decay width
\begin{eqnarray}
\label{5}
\mathrm{Im}\Delta E_A^{(2)}=\lim_{\eta\rightarrow 0}\eta \mathrm{Re}\langle A|\hat{S}_{\eta}^{(2)}|A\rangle.
\end{eqnarray}

Arranging all the terms of fourth order, which describe the pure two-photon decay width including a part of the radiative (one-loop) corrections to the one-photon width, one obtains 
\begin{eqnarray}
\label{6}
\mathrm{Im}\Delta E_A^{(4)} = \lim_{\eta\rightarrow 0} \eta \left[2 \mathrm{Re}\langle A|\hat{S}_{\eta}^{(4)}|A\rangle +
\right.
\\
\left.
\nonumber
\left|\langle A|\hat{S}_{\eta}^{(2)}|A\rangle \right|^2 - 2\left(\mathrm{Re}\langle A|\hat{S}_{\eta}^{(2)}|A\rangle \right)^2\right]\, ,
\end{eqnarray}
where the last two terms result from the expression $\langle A|\hat{S}_{\eta}^{(2)}|A\rangle^2$.

The total width $\Gamma_A$ of an excited electron state $A$ (specifying the initial state as $|A, 0_\gamma\rangle \equiv |A\rangle$) should follow (by definition) from the imaginary part of the total energy-shift: 
\begin{eqnarray}
\label{7}
\Gamma_A = -2 \mathrm{Im}\Delta E_A.
\end{eqnarray}
Respectively, after expansion of $\Delta E_A$ (up to order $e^4$) as
\begin{eqnarray}
\label{8}
\Gamma_A = -\lim_{\eta\rightarrow 0} 2\eta \left[
\mathrm{Re} \langle A|\hat{S}_{\eta}^{(2)}|A\rangle  + 
2 \mathrm{Re}\langle A|\hat{S}_{\eta}^{(4)}|A\rangle 
\right.
\\
\left.
\nonumber
+ \left|\langle A|\hat{S}_{\eta}^{(2)}|A\rangle \right|^2 - 
2\left(\mathrm{Re}\langle A|\hat{S}_{\eta}^{(2)}|A\rangle \right)^2\right].
\end{eqnarray}

As indicated above the adiabatic $S$-matrix $\hat{S}_\eta$ arises after introduction of the adiabatic switching function $f(\eta) = e^{-\eta|t|}$ in the QED interaction Hamiltonian. 
Assuming that no dynamic excitation of the system takes place during switching on and off the interaction, the adiabatic $S$-matrix remains unitary \cite{FGS91,Berest}. 
Moreover, all observable quantities calculated on the basis of adiabatic approach should not  
depend on the specific form used for the adiabatic factor after the limiting process 
$\eta\rightarrow 0$ has been performed. Therefore, we will apply the 'optical theorem' relations, see details in \cite{LSP-sep}: 
\begin{eqnarray}
\label{9}
-2\mathrm{Re}\langle A|\hat{S}^{(2)}|A\rangle &=& \sum\limits_{F\neq A}\left|\langle F|\hat{S}^{(1)}|A\rangle \right|^2,
\\
\label{10}
-2\mathrm{Re}\langle A|\hat{S}^{(4)}|A\rangle &=& \left|\langle A|\hat{S}^{(2)}|A\rangle \right|^2 +
\\
\nonumber
\sum\limits_{F\neq A}\left|\langle F|\hat{S}^{(2)}|A\rangle \right|^2 &+& \sum\limits_{F\neq A}
2\mathrm{Re}\langle A|\hat{S}^{(1)}|F\rangle
\langle F|\hat{S}^{(3)}|A\rangle .
\end{eqnarray}
to the adiabatic formulas (\ref{5}), (\ref{6}) and (\ref{8}). 

Then for the excited state $A$ without photons for the pure one-photon width one can find
\begin{eqnarray}
\label{11}
\Gamma_A^{(1)}= 
\lim_{\eta\rightarrow 0} \eta\,\sum\limits_{F\neq A}\left|\langle F|\hat{S}^{(1)}_{\eta}|A\rangle\right|^2 
\end{eqnarray}
and for the  two-photon width 
\begin{eqnarray}
\label{12}
\Gamma_A^{(2)}= 
\lim_{\eta\rightarrow 0}\eta \left[ 2\sum\limits_{F\neq A}\left|\langle F|\hat{S}^{(2)}_{\eta}|A\rangle\right|^2
+4\left(\mathrm{Re}\langle A|\hat{S}^{(2)}_{\eta}|A\rangle\right)^2\right]\qquad
\end{eqnarray}
or, employing Eq. (\ref{9}),
\begin{eqnarray}
\label{13}
\Gamma_A^{(2)}=\lim_{\eta\rightarrow 0} \eta \left[
2\sum\limits_{F\neq A}\left|\langle F|\hat{S}^{(2)}_{\eta}|A\rangle\right|^2 +
\qquad\qquad
\right.
\\
\nonumber
\left.
2\sum\limits_{2_\gamma}\left|\langle A, 2_\gamma|\hat{S}^{(2)}_{\eta}|A\rangle\right|^2 
+
\left(\sum\limits_{F'\neq A}\left|\langle F'|\hat{S}^{(1)}_{\eta}
|A\rangle\right|^2\right)^2
 \right].
\end{eqnarray}

In Eq. (\ref{13}) we have to distinguish between the final states ($F$) and ($F'$) 
for two-photon and for the one-photon transitions, respectively. It is important that the term $\left|\langle A|\hat{S}^{(2)}_{\eta}|A\rangle\right|^2$ has cancelled out in Eq. (\ref{12}). The last but one term in (\ref{13}), corresponding to apparently nonphysical transition $A\rightarrow A+2\gamma$, but formally present in the sum over $F$ states, will indeed cancel out in the final expression. The notation $\sum\limits_{2_\gamma}$ means here the integration over the frequencies of two photons.

The remaining term up to order $e^4$ containing radiative-correction effects is
\begin{eqnarray}
\label{14}
\Gamma_A^{{\rm rad}} = \lim_{\eta\rightarrow 0}\eta \,\sum\limits_{F\neq A}\,
2 \mathrm{Re}\langle A|\hat{S}^{(1)}|F\rangle\langle F|\hat{S}^{(3)}|A\rangle.
\end{eqnarray}
In the next section we will evaluate the one-photon and one-photon plus thermal interaction correction decay widths using Eqs. (\ref{11}) and (\ref{14}).

\section{One-photon decay width}
\label{sectionB}
\subsection{One-photon emission}
Evaluation of the decay width $\Gamma^{(1)}_A$ by the formula Eq. (\ref{11}) can be performed in a conventional manner. First, we evaluate the matrix element $\langle A'|\hat{S}^{(1)}_{\eta}|A\rangle$ ($\langle A'|\equiv \langle A',\vec{k},\vec{e}|\equiv \langle A'|\, \langle \vec{k},\vec{e}|$) for the emission of the photon with momentum $\vec{k}$ and polarization $\vec{e}$. The corresponding adiabatic $S_{\eta}$-matrix element reads
\begin{eqnarray}
\label{15}
\langle A'|\hat{S}^{(1)}_{\eta}|A\rangle=-i\,e\int d^4x\bar{\psi}_{A'}(x)\gamma_{\mu}A^{*}_{\mu}(x)\psi_A(x)e^{-\eta |t|}.\qquad
\end{eqnarray}
Here $\psi_A(x) = \psi_A(\vec{r})e^{-i E_A t}$, $\psi_A(\vec{r})$ is the solution of the Dirac equation for the atomic electron, $E_A$ is the Dirac energy, $\bar{\psi}_{A'} = \psi_{A'}^\dagger \gamma_0$ is the Dirac conjugated wave function with $\psi_{A'}^{\dagger}$ being its Hermitian conjugate and $\gamma_{\mu} = (\gamma_0, \vec\gamma)$ are the Dirac matrices and $A_{\mu}(x)$ is the photon wave function (electromagnetic field potential) 
\begin{eqnarray}
A_{\mu}(x)=\sqrt{\frac{2\pi}{\omega}}e^{(\lambda)}_{\mu}e^{ik_{\mu}x^{\mu}}.
\end{eqnarray}
Here $e^{(\lambda)}_{\mu}$ are the components of the photon polarization 4-vector, $x_{\mu}$ is the space-time 4-vector, $k_{\mu}$ is the photon momentum 4-vector with the space vector $\vec{k}$ and photon frequency $\omega=|\vec{k}|$. Using the transversality condition $\gamma_{\mu}e^{(\lambda)}_{\mu}=\vec{e}\vec{\alpha}$ ($\vec{e}$ is a transverse  space vector of the photon polarization), the wave function for the emitted/absorbed real photon  takes the form:
\begin{eqnarray}
\label{16}
\vec{A}(x) = \sqrt{\frac{2\pi}{\omega}}\,\vec{e}e^{i(\vec{k}\vec{r}-\omega t)}
 \equiv \sqrt{\frac{2\pi}{\omega}}\,e^{-i\omega t}\,\vec{A}(\vec{k},\vec{r}).
\end{eqnarray}

Now the integration over the time variable in Eq. (\ref{15}) yields essentially a representation of the $\delta$-function 
\begin{eqnarray}
\label{17}
\int\limits_{-\infty}^{\infty}dt\,e^{i(E_{A'}-E_{A}+\omega)t-\eta |t|}=\frac{2\eta}{(\omega_{AA'}-\omega)^2+\eta^2} 
\\
\nonumber
\equiv 2\pi\,\delta_\eta (\omega_{AA'}-\omega),
\end{eqnarray}
where $\lim\limits_{\eta\rightarrow 0}\delta_{\eta}(x)=\delta (x)$.
As the next step we perform the integration over the photon frequency. Taking Eq. (\ref{17}) by square modulus, multiplying by $\omega$ and integrating, we obtain
\begin{eqnarray}
\label{18}
4\eta^2\int\limits_0^{\infty}\frac{\omega d\omega}{\left[(\omega_{AA'}-\omega)^2+\eta^2\right]^2}=
\\
\nonumber
4\eta^2\left[ 
\frac{\pi\omega_{AA'}}{4\eta^3} +\frac{1}{2\eta^2}+\frac{\omega_{AA'}}{2\eta^3}{\rm arctg}\left(\frac{\omega_{AA'}}{\eta}\right)
\right].
\end{eqnarray}
Having in mind the factor $\lim\limits_{\eta\rightarrow 0}\eta$ we can replace Eq. (\ref{18}) by
\begin{eqnarray}
\label{19}
4\eta^2\int\limits_0^{\infty}\frac{\omega d\omega}{\left[(\omega_{AA'}-\omega)^2+\eta^2\right]^2}=\frac{2\pi\omega_{AA'}}{\eta}.
\end{eqnarray}

Multiplying the square modulus of Eq. (\ref{15}) by the factor $d\vec{\nu}/(2\pi)^{3}$ (the phase volume) with the account for (\ref{18}) and with the summation over the electron states, lower by energy than the state $A$, we arrive at
\begin{eqnarray}
\label{20}
\Gamma_A^{(1)}=\frac{e^2}{2\pi}
\sum\limits_{A'}^{E_{A'}<E_A} \omega_{AA'} \sum\limits_{\vec{e}}\int d\vec{\nu} 
\left|\left( \vec{\alpha}\vec{A}^* \right)_{A'A} \right|^2,\qquad
\end{eqnarray}
where $\vec{\alpha}$ is the Dirac matrix. In the derivation above the manipulations with $\Delta$-functions, like in 'normal' $S$-matrix formalism, are avoided. Multiplying the result by the adiabatic parameter $\eta$ in Eq. (\ref{11}) plays the same role as dividing the result by the time $T$: the adiabatic factor $\eta$ has the dimensionality s$^{-1}$. Note, that in this approach the automatic exclusion of the transitions to the states higher than $A$ in the summation over $F$ in Eq. (\ref{11}) does not occur and we have to refer to the energy conservation law.

In the nonrelativistic limit and electric dipole approximation Eq. (\ref{20}) after the integration over photon emission direction, summation over photon polarizations, summation over projections of final states and averaging over projections of initial state Eq. (\ref{20}) takes the form
\begin{eqnarray}
\Gamma_A^{(1)}=\sum\limits_{A'}^{E_{A'}<E_A}W_{AA'},
\end{eqnarray}
where $W_{AA'}$ is the partial transition rate
\begin{eqnarray}
\label{10a}
W_{AA'}=\frac{e^2}{2l_{A}+1}\sum\limits_{m_{A}m_{A'}}\frac{4}{3}\omega_{AA'}^3|\langle A'| \vec{r}| A\rangle|^2.
\end{eqnarray}
Below we will be also interested in BBR-induced one-photon transition rate which is given by
\begin{eqnarray}
\label{10b}
W^{{\rm ind}}_{AA'}=\frac{e^2}{2l_{A}+1}\sum\limits_{m_{A}m_{A'}}\frac{4}{3}\omega_{AA'}^3|\langle A'| \vec{r}| A\rangle|^2 
n_{\beta}(\omega_{AA'}),\qquad
\end{eqnarray}
where $n_{\beta}(\omega)=(e^{\omega \beta}-1)^{-1}$, $\beta=(k_{B}T)^{-1}$, $k_{B}$ is the Boltzman constant and $T$ is the radiation temperature in kelvin.

\subsection{One-photon emission with vertex correction}

\begin{figure}[hbtp]
\centering
\includegraphics[scale=0.2]{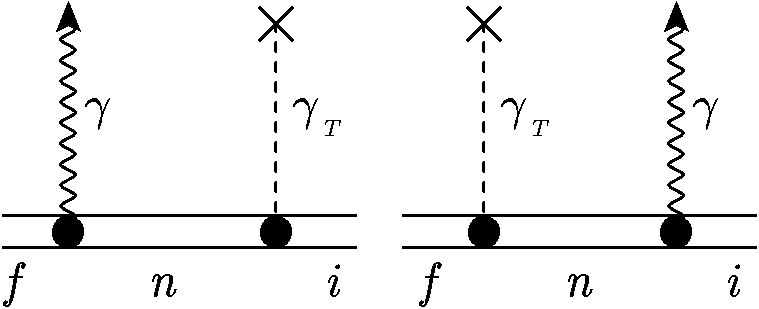}
\caption{Feynman diagrams representing the thermal correction on the thermal interaction potential. A wavy line ($\gamma$) indicates the photon emission process, a dashed line ($\gamma_T$) corresponds to the thermal Coulomb photon exchange of a bound electron with a nucleus. The double solid line denotes the bound electron in the nucleus field (the Furry picture). Notations $i$ and $f$ represent the initial and final states of a bound electron, respectively, and $n$ corresponds to the intermediate state represented in the electron propagator.}
\label{fig1}
\end{figure}

Recently, the QED approach at finite temperature has been used to describe the Feynman diagram corresponding to the photon exchange between a bound electron and a nucleus \cite{S-2020,SZA-2020}. As a consequence, a thermal interaction potential was found that corresponds to the Coulomb part of the thermal photon propagator. The lowest-order thermal correction to the level energies arising from the series expansion of this potential at room temperature is of the order of accuracy in modern experiments \cite{MHH}. Here we consider this thermal correction to the one-photon emission process. The inclusion of thermal interaction in the process of one-photon emission is schematically depicted by the Feynman graphs in Fig.~\ref{fig1}.

The corresponding $S_\eta$-matrix element for the graph a) reads
\begin{eqnarray}
\label{21}
S^{(3)}_{\eta,a} = (-i e)^2 i Ze \int d^4x_1 d^4x_2 d^4x_3 \bar{\psi}_f(x_1)\gamma^\mu A_\mu(x_1) \qquad
\\
\nonumber
\times
e^{-\eta |t_1|}S(x_1,x_2)
e^{-\eta |t_2|} \gamma^\nu 
D^\beta_{\nu\lambda}(x_2,x_3)
 j^\lambda(x_3)\psi_i(x_2),\qquad
\end{eqnarray}
where $Ze$ is the external charge (nucleus), $S(x_1,x_2)$ is the electron propagator and 
$j^\lambda(x)$ is the external nuclear current which can be written using the Fourier transform as
\begin{eqnarray}
\label{current}
j^\lambda(x) =\int\frac{d^4q}{(2\pi)^4}e^{iqx}j^{\lambda}(q).
\end{eqnarray}
The eigenmode decomposition of $S(x_1,x_2)$ with respect to one-electron eigenstates is
\begin{eqnarray}
\label{22}
S(x_1,x_2) = \frac{i}{2\pi}\int\limits_{-\infty}^\infty d\Omega\, e^{-i\Omega(t_1-t_2)}
\sum\limits_n\frac{\psi_n(\vec{r}_1)\bar{\psi}_n(\vec{r}_2)}{\Omega-E_n(1-i0)},\qquad
\end{eqnarray}
where summation runs over the entire Dirac spectrum.

The thermal interaction shown by the dashed line in Fig.~\ref{fig1} corresponds to the zero component of the thermal photon propagator, $D^\beta_{\nu\lambda}(x_2,x_3)$. In the case of an infinitely heavy and static external charge (nucleus), it is convenient to use the contour integral representation found in \cite{S-2020}:
\begin{eqnarray}
\label{23}
D_{\mu\nu}^\beta(x,x') = -4\pi i g_{\mu\nu}\int\limits_{C_1}\frac{d^4k}{(2\pi)^4}\frac{e^{ik(x-x')}}{k^2}n_\beta(|\vec{k}|),\qquad
\end{eqnarray}
where $g_{\mu\nu}$ is the metric tensor, $k^2\equiv k_0^2-|\vec{k}|^2$.
Substitution of Eq. (\ref{23}) into (\ref{21}) allows the integration over $x_3$ that leads to the Fourier transform of the external current $j^0(x_3)$. Then
\begin{eqnarray}
\label{24}
S^{(3)}_{\eta,a} = -i Ze^3 \int d^4x_1 d^4x_2\bar{\psi}_f(x_1)\gamma^\mu A_\mu(x_1)e^{-\eta |t_1|}  \times\qquad
\\
\nonumber
S(x_1,x_2) e^{-\eta |t_2|} (-4\pi i) \int\limits_{C_1}\frac{d^4q}{(2\pi)^4}\frac{e^{iqx_2}}{q^2}n_\beta(|\vec{q}|) j^0(q)\psi_i(x_2).\qquad
\end{eqnarray}

Employing the static limit to the nuclear current $j^0(q)=2\pi\delta(q_0)\rho(\vec{q})\approx 2\pi\delta(q_0)$ (in approximation of point-like nucleus), where $\delta(q_0)$ is the Dirac $\delta$-function, one can find
\begin{eqnarray}
\label{25}
S^{(3)}_{\eta,a} = -4\pi Ze^3 \int d^4x_1 d^4x_2\bar{\psi}_f(x_1)\gamma^\mu A_\mu(x_1)e^{-\eta |t_1|}  
\\
\nonumber
\times
S(x_1,x_2) e^{-\eta |t_2|} \int\frac{d^3q}{(2\pi)^3}\frac{e^{i\vec{q}\vec{r}_2}}{\vec{q}^2}n_\beta(|\vec{q}|)\psi_i(x_2)
.
\end{eqnarray}
It should be noted here that the same expression could immediately be written in the thermal Coulomb gauge, see \cite{S-2020,SZA-2020}.

The following evaluation corresponds to the integration over the time variables $t_1$ and $t_2$. Substituting Eq. (\ref{22}) into (\ref{25}) with the use of Eqs. (\ref{16}) and (\ref{17}), we arrive at
\begin{gather}
\label{26}
S^{(3)}_{\eta,a} = -2 i Ze^3 \int d^3r_1 d^3r_2 \bar{\psi}_f(\vec{r}_1)
\sqrt{\frac{2\pi}{\omega}}
(\vec{\alpha}_1\vec{A}(\vec{k},\vec{r}_1))\times
\\
\nonumber
\sum\limits_n \psi_n(\vec{r_1})\bar{\psi}_n(\vec{r}_2)
\int\frac{d^3q}{(2\pi)^3}n_\beta(|\vec{q}|)\frac{e^{i\vec{q}\vec{r}_2}}{\vec{q}^2}\psi_i(\vec{r}_2)\times
\\
\nonumber
\frac{4\pi \eta}{(E_i-E_f-\omega)^2+4\eta^2}
\frac{E_f+E_i-2E_n+\omega +4i\eta}{(E_i-E_n+i\eta)(E_f-E_n+\omega+i\eta)}
,
\end{gather}
where the relation $\psi_a(x)=\psi_a(\vec{r})e^{-iE_a t}$ and integration over the frequency $\Omega$, see Eq.(\ref{22}), were employed.
Repeating all the calculations above, for the graph b) in Fig.~\ref{fig1} one can find
\begin{eqnarray}
\label{27}
S^{(3)}_{\eta,b} = -2 i Ze^3 \int d^3r_1 d^3r_2 \bar{\psi}_f(\vec{r}_1)\int\frac{d^3q}{(2\pi)^3}n_\beta(|\vec{k}|)\times\qquad
\nonumber
\\
\frac{e^{i\vec{q}\vec{r}_1}}{\vec{q}^2}\sum\limits_n \psi_n(\vec{r_1})\bar{\psi}_n(\vec{r}_2)
\sqrt{\frac{2\pi}{\omega}}(\vec{\alpha}_2\vec{A}(\vec{k},\vec{r}_2))\psi_i(\vec{r}_2)\times\qquad
\\
\nonumber
\frac{4\pi \eta}{(E_i -E_f -\omega)^2+4\eta^2}
\frac{E_f+E_i-2E_n-\omega +4i\eta}{(E_f-E_n+i\eta)(E_i-E_n-\omega+i\eta)}.
\end{eqnarray}

To determine the radiative correction to the transition probability, it is also necessary to calculate $S^{(1)}_\eta$:
\begin{eqnarray}
\label{28}
S^{(1)}_\eta = \frac{-ie \,2\eta \sqrt{\frac{2\pi}{\omega}}}{(E_i-E_f-\omega)^2+\eta^2}
\int d^3r \bar{\psi}_f(\vec{r})(\vec{\alpha}\vec{A}(\vec{k},\vec{r}))\psi_i(\vec{r}).\qquad
\end{eqnarray}
Then, according to Eq. (\ref{14}), 
the summary contribution is
\begin{eqnarray}
\label{29}
\Delta W_{if}^{{\rm rad}} = \frac{2^6\pi^2 Ze^4}{\omega}\frac{d^3k}{(2\pi)^3}\mathrm{Re}\langle i | \vec{\alpha}\vec{A} | f \rangle\times \qquad
\\
\nonumber
\lim\limits_{\eta\rightarrow 0}\frac{\eta^3}{((E_i-E_f-\omega)^2+\eta^2)((E_i-E_f-\omega)^2+4\eta^2)}
\\
\nonumber
\times
\left[
 \sum\limits_n \frac{\langle f | \vec{\alpha}\vec{A}^* | n \rangle \langle n | V^\beta(\vec{r}) | i \rangle (E_f+E_i-2E_n+\omega +4i \eta)}{(E_i-E_n+ i \eta)(E_f-E_n+\omega+ i \eta)} 
 \right.
 \\\nonumber
 \left.
 + \sum\limits_n \frac{\langle f | V^\beta(\vec{r}) | n \rangle \langle n | \vec{\alpha}\vec{A}^* | i \rangle (E_f+E_i-2E_n-\omega+4i \eta)}{(E_f-E_n+ i \eta)(E_i-E_n-\omega+ i \eta)}
 \right],
\end{eqnarray}
where we used notation Eq. (\ref{16}) and definition 
\begin{eqnarray}
V^\beta(\vec{r})\equiv\int\frac{d^3q}{(2\pi)^3}n_\beta(|\vec{q}|)\frac{e^{i\vec{q}\vec{r}}}{\vec{q}^2}.
\end{eqnarray}
In Eq. (\ref{29}) the phase volume of the emitted photon is inserted also.

Finally, one should integrate over the frequency $\omega$. Usage $d^3k = \omega^2 d\omega d\vec{\nu}$ results in
\begin{widetext}
\begin{eqnarray}
\label{30}
\Delta W_{if}^{{\rm rad}} = \frac{8 Ze^4}{3}  \lim\limits_{\eta\rightarrow 0}\mathrm{Re}\langle i | \vec{\alpha}\vec{A} | f \rangle \left[
\nonumber
\sum\limits_n \frac{\langle f | \vec{\alpha}\vec{A}^* | n \rangle \langle n | V^\beta(\vec{r}) | i \rangle [-(E_f-E_i)(E_i-E_n)- 3i (E_f-E_i)\eta - \eta^2]}{(E_i-E_n+ i \eta)(E_i-E_n + 2i \eta)}
\right.
\\
\left.
+ \sum\limits_n \frac{\langle f | V^\beta(\vec{r}) | n \rangle \langle n | \vec{\alpha}\vec{A}^* | i \rangle [-(E_f-E_i)(E_f-E_n)-3 i (E_f-E_i)\eta + \eta^2]}{(E_f-E_n+ i \eta)(E_f- E_n + 2 i\eta) }\right]
d\vec{\nu}
.   \qquad
\end{eqnarray}
\end{widetext}

The contributions to the thermal radiative correction, Eq. (\ref{30}), can be considered as reducible and irreducible parts. The first one corresponds to the case of $n=i$ and $n=f$ in the first and second terms, and the second, irreducible contribution, is represented by $n\neq i$ and $n\neq f$, respectively. The calculations are not difficult, but in the presence of poles in the energy denominators for the reducible part, one should at first to use the series expansion over the vanishing energy difference. Taking into account the limit $\eta\rightarrow 0$ and that the product of operators in numerators are purely real, the result reforms to
\begin{gather}
\label{31}
\nonumber
\Delta W_{if}^{{\rm rad}} = \frac{8 Ze^4}{3}\langle i|\vec{\alpha}\vec{A}|f\rangle  
\left[ \omega_{if} \mathop{{\sum}'}
\limits_n \frac{\langle f | \vec{\alpha}\vec{A}^* | n \rangle\langle n | V^\beta(\vec{r}) | i \rangle}{E_i-E_n}
\right.
\\\nonumber
 + \omega_{if} \mathop{{\sum}'}\limits_n \frac{\langle f | V^\beta(\vec{r}) | n \rangle\langle n | \vec{\alpha}\vec{A}^* | i \rangle}{E_f-E_n} 
+ \frac{1}{2}\langle f | \vec{\alpha}\vec{A}^* | i \rangle
\\
\times
\langle i | V^\beta(\vec{r}) | i \rangle 
\left.
- \frac{1}{2}\langle f | V^\beta(\vec{r}) | f \rangle\langle f | \vec{\alpha}\vec{A}^* | i \rangle \right]
d\vec{\nu}
,\qquad
\end{gather}
where $\omega_{if}\equiv E_i-E_f$, the prime of the sum sign means the absence of corresponding state in the summation over the entire spectrum.
%

Now we turn to the thermal interaction potential. In \cite{S-2020} it was found that the matrix element $\langle A | V^\beta(\vec{r}) | A\rangle$ diverges. To avoid infrared divergence of this type, a regularization procedure can be applied, see \cite{S-2020}. However, this procedure is redundant in this case (without affecting the result). As it was found in \cite{ZSL-1ph,ZAS-2l} for the thermal self-energy radiative corrections to one- and two-photon decay rates, all infrared divergences vanish for the complete set of the same order diagrams. For vertex-type corrections, this divergence can also be singled out explicitly. Integration of $V^\beta$ over angles in momentum space yields
\begin{eqnarray}
\label{33}
V^\beta(r) = \frac{1}{2\pi^2}\int\limits_0^\infty dq\, n_\beta(q)\frac{\sin qr}{qr}
\\
\nonumber
\approx \frac{1}{2\pi^2}\int\limits_0^\infty dq\, \frac{1-\frac{1}{6}q^2 r^2}{e^{\beta q}-1}.
\end{eqnarray}
Here we have used that the Planck distribution function cuts off the high-frequency range at relevant temperatures and $r$ is limited by the atomic radius. The first term of Eq. (\ref{33}) represents a divergent contribution which, however, turns to be zero in view of the orthogonality of wave functions in the irreducible part of Eq. (\ref{32}). For the reducible contributions (last two terms in Eq. (\ref{32})) divergent part of Eq. (\ref{33}) cancels out due to the opposite signs. Thus, there is no divergence in Eq. (\ref{32}), and the dominant contribution can be reduced to the expression:
\begin{eqnarray}
\label{34}
\Delta W_{if}^{{\rm rad}} = -\frac{4 Ze^4\zeta(3)}{9\pi^2\beta^3}\langle i|\vec{\alpha}\vec{A}|f\rangle \times  \qquad
\\
\nonumber
\left[ 
\omega_{if} 
\mathop{{\sum}'}\limits_n \frac{\langle f | \vec{\alpha}\vec{A}^* | n \rangle\langle n | r^2 | i \rangle}{E_i-E_n}
 + 
\omega_{if}  
 \mathop{{\sum}'}\limits_n \frac{\langle f | r^2 | n \rangle\langle n | \vec{\alpha}\vec{A}^* | i \rangle}{E_f-E_n} 
\right.
\\
\left.
\nonumber
+ \frac{1}{2}\langle f | \vec{\alpha}\vec{A}^* | i \rangle\langle i | r^2| i \rangle - \frac{1}{2}\langle f | r^2 | f \rangle\langle f | \vec{\alpha}\vec{A}^* | i \rangle \right]
d\vec{\nu}
,\qquad
\end{eqnarray}
where the second term in Eq. (\ref{33}) was integrated over $q$ and $\zeta$ is the Riemann zeta function. It is important to note that the exact cancellation of infrared divergences in the equations  was also found in \cite{ZSL-1ph,ZAS-2l}.  

For the relativistic calculations the following decomposition for emission operator can be used \cite{drake, ZSLP-report}
\begin{eqnarray}
\vec{\alpha}\vec{A}^{*}=\sum\limits_{\lambda L M}\left[ \vec{e}^{\;*}\vec{Y}_{LM}^{(\lambda)}(\vec{\nu})\right]\widetilde{a}^{(\lambda)*}_{LM}(r),
\end{eqnarray}
where $L$ and $M$ is the photon angular momentum and its projection. Terms with $\lambda = 1$ are electric multipoles and terms with $\lambda = 0$ are magnetic multipoles. Then the reduction of matrix elements can be performed with the use of relation 
\begin{eqnarray}
\langle n'l'j'm' | \widetilde{a}^{(\lambda)*}_{LM} | nljm \rangle = (-1)^{j'-m'}
\\\nonumber
\times
\begin{pmatrix}
j' & L & j \\
-m' & M & m
\end{pmatrix}
(-i)^{L+\lambda -1}(-1)^{j'-1/2}
\\\nonumber
\times
\sqrt{\frac{4\pi}{2L+1}}[j',j]^{1/2}
\begin{pmatrix}
j' & L & j \\
1/2    & 0 & -1/2
\end{pmatrix}
\overline{M}^{(\lambda, L)}_{n'n}
,
\end{eqnarray}
where 
\begin{eqnarray}
\label{23y}
\overline{M}^{(1,L)}_{n'l'nl}=\left(\frac{L}{L+1}\right)^{1/2}\left[\left(\kappa'-\kappa\right)I^{+}_{L+1}
\right.
\\\nonumber
+
\left.
(L+1)I^{-}_{L+1}\right]-\left(\frac{L+1}{L}\right)^{1/2}
\\\nonumber
\times
\left[\left(\kappa'-\kappa\right)I^{+}_{L-1}-LI^{-}_{L-1}\right]
\\
\nonumber
 - G\left[(2L+1)J_{L}+\left(\kappa'-\kappa\right)
\right.
\\\nonumber
\left. 
\times
\left(I^{+}_{L+1}-I^{+}_{L-1}\right)-LI^{-}_{L-1}+\left(L+1\right)I^{-}_{L+1}\right],
\end{eqnarray}
\begin{eqnarray}
\label{24y}
\overline{M}^{(0,L)}_{n'l'nl}=\frac{2L+1}{\left[L(L+1)\right]^{1/2}}\left(\kappa'+\kappa\right)I^{+}_{L},
\end{eqnarray}
\begin{eqnarray}
\label{24-0y}
\overline{M}^{(-1,L)}_{n'l'nl}=
-G_{L}\left[(2L+1)J_{L}
+\left(\kappa'-\kappa\right)
\right.
\\\nonumber
\left.
\times
\left(I^{+}_{L+1}-I^{+}_{L-1}\right)
-LI^{-}_{L-1}+\left(L+1\right)I^{-}_{L+1}\right],
\end{eqnarray}
and $I^{\pm}_{L}$,  $J_{L}$ are the radial integrals
\begin{eqnarray}
\label{25y}
I^{\pm}_{L}=\int\limits_{0}^{\infty}\left(g_{n'l'j'}f_{nlj}\pm f_{n'l'j'}g_{nlj}\right)j_{L}\left(\omega r\right)dr,
\end{eqnarray}
\begin{eqnarray}
\label{26y}
J_{L}=\int\limits_{0}^{\infty}\left(g_{n'l'j'}g_{nlj}+ f_{n'l'j'}f_{nlj}\right)j_{L}\left(\omega r\right)dr.
\end{eqnarray}
Here $g_{n'l'j'}$ and $f_{n'l'j'}$ are the large and small components of the radial Dirac wave function, $\kappa$ is the Dirac angular number, $G_{L}$ is the gauge parameter (the 'length form' $G_{L}= \sqrt{(L+1)/L}$ is used for the numerical calculations in this paper) and $j_{L}$ is the spherical Bessel function. The matrix element of thermal potential can be found with the use of standard angular algebra
\begin{eqnarray}
\langle n'l'j'm'| r^2  | nljm \rangle
=
\delta_{l'l}\delta_{j'j}\delta_{m'm}
\\\nonumber
\times
\int\limits_{0}^{\infty}\left(g_{n'l'j'}g_{nlj}+ f_{n'l'j'}f_{nlj}\right)r^2 dr,
\end{eqnarray}

Integration over angles and summation over polarizations can be performed with the orthogonality condition
\begin{eqnarray}
\label{summation}
\sum\limits_{\vec{e}}\int d\vec{\nu}\left[ \vec{e}^{\;*}\vec{Y}_{L'M'}^{(\lambda')}(\vec{\nu})\right]^*\left[ \vec{e}^{\;*}\vec{Y}_{LM}^{(\lambda)}(\vec{\nu})\right]=
\\
\nonumber
= \delta_{L'L}\delta_{M'M}\delta_{\lambda'\lambda}.
\end{eqnarray}
Then the total radiative correction to the one-photon transition, after summation over photon polarization and integration over photon emission direction with the use of Eq. (\ref{summation}), summation over the projections of the final state and averaging over projection of the initial state, is
\begin{eqnarray}
\label{relmain}
\Delta W_{if}^{{\rm rad}} = -\frac{1}{2j_i+1}\frac{4 Ze^4\zeta(3)}{9\pi^2\beta^3} \sum\limits_{\lambda LM}\sum\limits_{m_{i}m_{f}}\langle i|\widetilde{a}^{(\lambda)}_{LM}f\rangle  
\times \qquad
\\
\nonumber
\left[ 
\omega_{if}  
\mathop{{\sum}'}\limits_n \frac{\langle f | \widetilde{a}^{(\lambda)*}_{LM}| n \rangle\langle n | r^2 | i \rangle}{E_i-E_n}
 + 
\omega_{if}   
 \mathop{{\sum}'}\limits_n \frac{\langle f | r^2 | n \rangle\langle n | \widetilde{a}^{(\lambda)*}_{LM} | i \rangle}{E_f-E_n} 
\right.
\\
\left.
\nonumber
+ \frac{1}{2}\langle f | \widetilde{a}^{(\lambda)*}_{LM} | i \rangle\langle i | r^2| i \rangle - \frac{1}{2}\langle f | r^2 | f \rangle\langle f | \widetilde{a}^{(\lambda)*}_{LM} | i \rangle \right]
d\vec{\nu}.
\end{eqnarray}

The results of evaluation of thermal correction, Eq. (\ref{relmain}), to $2p_{1/2}\rightarrow 1s_{1/2}+\gamma(\mathrm{E1})$ and $2p_{1/2}\rightarrow 1s_{1/2}+\gamma(\mathrm{E1})$ transitions in H-like ions with different nuclear charge $Z$ are given in Table \ref{tab:2}.
\begin{center}
\begin{table}
\caption{Numerical values of electric dipole decay rates  $W_{if}$ and thermal correction $ \Delta W^{{\rm rad}}_{if}$, Eq. (\ref{relmain}), for $2p_{1/2}\rightarrow 1s_{1/2} + \gamma({\rm E1})$ (upper line) and $5p_{1/2}\rightarrow 1s_{1/2} + \gamma({\rm E1})$ (lower line) transitions in H-like ions at room temperature. The first column shows the nuclear charge $Z$. Calculations are performed within the fully relativistic approach. All values are given in s$^{-1}$.}
\begin{tabular}{l c r}
\hline
\hline
$Z$  & $W_{if}$ & $ \Delta W^{{\rm rad}}_{if}$, Eq. (\ref{relmain}) \\
\hline
1 & $ 6.26835\times 10^{8} $ & $ 1.463\times 10^{-5}$ \\
  & $ 3.43942\times 10^{7} $ & $ 2.821\times 10^{-4}$ \\

2 & $1.00295\times 10^{10}$ & $ 2.925\times 10^{-5}$ \\
  & $5.50276 \times 10^8$   & $ 5.642\times 10^{-4}$ \\

5 & $3.91828\times 10^{11}$ & $  7.305\times 10^{-5}$ \\
  & $2.14864\times 10^{10}$ & $  1.409\times 10^{-3}$ \\

10  & $6.27219\times 10^{12}$ & $ 1.455\times 10^{-4}$ \\
    & $3.43281\times 10^{11}$ & $ 2.808\times 10^{-3}$ \\

20 & $1.00545\times 10^{14}$ & $  2.859\times 10^{-4}$ \\
   & $5.46004\times 10^{12}$ & $  5.533\times 10^{-3}$ \\
  
35 & $9.47919\times 10^{14}$ & $  4.764\times 10^{-4}$ \\
   & $5.03405\times 10^{13}$ & $  9.281\times 10^{-3}$ \\

50 & $3.98002\times 10^{15}$ & $  6.280\times 10^{-4}$ \\
   & $2.03662\times 10^{14}$ & $  1.236\times 10^{-2}$ \\

70 & $1.55211\times 10^{16}$ & $   7.419\times 10^{-4}$ \\
   & $7.32772\times 10^{14}$  & $  1.486\times 10^{-2}$ \\

92 & $4.72597\times 10^{16}$ & $   7.122\times 10^{-4}$ \\
   & $1.90889\times 10^{15}$ & $   1.422\times 10^{-2}$ \\
\hline
\hline
\end{tabular}
\label{tab:2}
\end{table}
\end{center}
From Tables ~\ref{tab:2} it follows that the thermal correction Eq. (\ref{relmain}) is much less than the corresponding one-photon electric dipole transition. Nonetheless, according to the parametric estimation ($\sim \alpha^6$ for the hydrogen atom) it can be compared with two-photon E1M1 and E1E2 transitions, see \cite{LabShonSol}. In particular, the sum of the two-photon transition rates for hydrogen is $W^{\rm E1M1+E1E2}_{2p-1s} = 1.627\times 10^{-5}$ s$^{-1}$, while the vertex-type thermal correction is $ 1.462\times 10^{-5}$ s$^{-1}$. In turn, for a H-like ion with $ Z=2 $, the total two-photon transition rate is $W^{\rm E1M1+E1E2}_{2p-1s} = 4.164\times 10^{-3}$ s$^{-1}$ \cite{LabShonSol}, and the thermal correction is $ 2.925\times 10^{-5}$ s$^{-1}$. This tendency persists at high values of the nuclear charge $Z$. Notwithstanding, the rise in temperature should lead to a cubic increase of thermal correction, which makes it essential for Ly$_\alpha$ transitions in light one-electron ions.

The thermal correction, Eq. (\ref{relmain}), calculated for the magnetic dipole $2s_{1/2}\rightarrow 1s_{1/2}+\gamma(\mathrm{M1})$ transition in different H-like ions is collected in Table~\ref{tab:3}.
\begin{center}
\begin{table}
\caption{Numerical values of the magnetic dipole transition rate $W_{2s1s}$ and thermal correction $\Delta W_{2s1s}^{{\rm rad}}$ for different H-like ions at room temperature. Calculations are performed within the fully relativistic approach. The first column shows the nuclear charge $Z$. All values are given in s$^{-1}$. }
\begin{tabular}{l c c}
\hline
\hline
$Z$  & $W_{2s1s} $ & $ \Delta W_{2s1s}^{{\rm rad}} $   \\
\hline
1 & $2.4959\times 10^{-6}$ & $ 3.023\times 10^{-16}$ \\

2 & $2.5563\times 10^{-3}$ & $ 9.683\times 10^{-15}$ \\

5 & $24.4076$ & $ 9.528\times 10^{-13}$ \\

10  & $2.5100\times 10^4$ & $ 3.131\times 10^{-11}$ \\

20 & $2.6149\times 10^7$ & $  1.107\times 10^{-9}$ \\
  
35 & $7.3930\times 10^9$ & $ 2.293\times 10^{-8}$ \\

50 & $2.8291\times 10^{11}$ & $ 1.804\times 10^{-7}$ \\

70 & $9.5829\times 10^{12}$ & $ 1.418\times 10^{-6}$ \\

92 & $1.9241\times 10^{14}$ & $ 8.148\times 10^{-6}$ \\

\hline
\hline
\end{tabular}
\label{tab:3}
\end{table}
\end{center}


Within the framework of the dipole approximation and the nonrelativistic limit $kr\ll 1$, the expression (\ref{relmain}) can be significantly simplified. In this case, the emission operator $\vec{\alpha}\vec{A}$, see Eq. (\ref{16}), is reduced to $(\vec{\alpha}\vec{e})$. Then the summation over photon polarization with integration over the photon emission direction, the summation over the projections of the final state, and averaging over projection of the initial state, give
\begin{eqnarray}
\label{32}
\Delta  W_{if}^{{\rm rad}} = -\frac{1}{2l_i+1}  \frac{2^5Ze^4\zeta(3)}{27\pi\beta^3} \omega_{if}^2\sum\limits_{m_{i}m_{f}}  \langle i|\vec{r}|f \rangle \qquad
\\
\nonumber
\times
 \left[
\mathop{{\sum}'}\limits_n \frac{\omega_{fn}\langle f | \vec{r} | n \rangle\langle n | r^2  | i \rangle}{E_i-E_n} 
+  
\mathop{{\sum}'}\limits_n \frac{ \omega_{ni}\langle f | r^2 | n \rangle\langle n | \vec{r} | i \rangle  }{E_f-E_n}
\right.
\\
\nonumber
 + 
\frac{1}{2}\langle f | \vec{r} | i \rangle\langle i | r^2 | i \rangle
\left.
 - 
\frac{1}{2}\langle f | r^2 | f \rangle 
\langle f | \vec{r} | i \rangle\right],
\end{eqnarray}
where the relation $\langle a |\vec{p}|b\rangle = i\omega_{ab} \langle a |\vec{r}|b\rangle$ was used. The numerical results of Eq. (\ref{32}) for various dipole one-photon $i\rightarrow f$ transitions in the hydrogen atom at room temperature are collected in Table~\ref{tab:1}. Evaluation at other temperatures can be easily obtained by multiplying by the coefficient $(T/300)^3$. 
\begin{widetext}
\begin{center}
\begin{table}
\caption{Numerical values of the thermal corrections $\Delta W_{if}^{{\rm rad}}$ and $\Delta W_{if}^{{\rm v}}$ for different one-photon transitions at room temperature ($ T=300 $ K) in the hydrogen atom. The first column shows the considered one-photon transitions. In the second column the natural one-photon transition rates $W_{if}$ are calculated within the dipole and nonrelativistic approximations. The third and fourth columns contain vertex-type thermal correction $\Delta  W_{if}^{{\rm rad}}$ and $\Delta W_{if}^{{\rm v}}$, respectively. In the fifth column, the total contribution $\Delta W_{if}^{{\rm total}} = \Delta  W_{if}^{{\rm rad}}+\Delta W_{if}^{{\rm v}}$ is presented for the corresponding transitions. All values are in s$ ^{-1} $. }
\begin{tabular}{l c c c c}
\hline
\hline
Transition & $W_{if}$, Eq. (\ref{10a}) & $\Delta  W_{if}^{{\rm rad}}$, Eq. (\ref{32}) & $\Delta  W_{if}^{{\rm v}}$, Eq. (\ref{29b}) & $\Delta W_{if}^{{\rm total}}$\\
\hline
2p-1s & $6.268\times 10^8$ & $ 1.463\times 10^{-5} $ & $ 2.285\times 10^{-5}$ & $  3.748\times 10^{-5}
 $\\

3p-1s & $1.673\times 10^8$ & $ 5.562\times 10^{-5} $ & $ 3.394\times 10^{-5}$ & $  8.956\times 10^{-5}
 $\\
 
3p-2s & $2.246\times 10^7$ & $ 6.147\times 10^{-6} $ & $ 2.277\times 10^{-5}$ & $  2.892\times 10^{-5}
 $\\

4p-1s & $6.822\times 10^7$ & $ 1.405\times 10^{-4} $ & $ 4.429\times 10^{-5}$ & $  1.847\times 10^{-4}
 $\\

4p-2s & $9.673\times 10^6$ & $ 1.595\times 10^{-5} $ & $ 2.937\times 10^{-5}$ & $  4.532\times 10^{-5}
 $\\

4p-3s & $3.067\times 10^6$ & $ 7.475\times 10^{-6} $ & $ 2.529\times 10^{-5}$ & $  3.277\times 10^{-5}
 $\\

5p-1s & $3.439\times 10^7$ & $ 2.821\times 10^{-4} $ & $ 5.469\times 10^{-5}$ & $  3.368\times 10^{-4}
 $\\

5p-2s & $4.951\times 10^6$ & $ 3.536\times 10^{-5} $ & $ 3.506\times 10^{-5}$ & $  7.042\times 10^{-5}
 $\\

5p-3s & $1.639\times 10^6$ & $ 8.787\times 10^{-6} $ & $ 3.039\times 10^{-5}$ & $  3.918\times 10^{-5}
 $\\

5p-4s & $7.376\times 10^5$ & $ 7.340\times 10^{-6} $ & $ 2.849\times 10^{-5}$ & $  3.583\times 10^{-5}
 $\\
\hline
\hline
\end{tabular}
\label{tab:1}
\end{table}
\end{center}
\end{widetext}

\subsection{Vertex contribution to the transition frequency}

\begin{figure}[hbtp]
\caption{Vertex correction to the atomic energy level $i$. All designations are the same as in Fig.~\ref{fig1}.}
\centering
\includegraphics[scale=0.15]{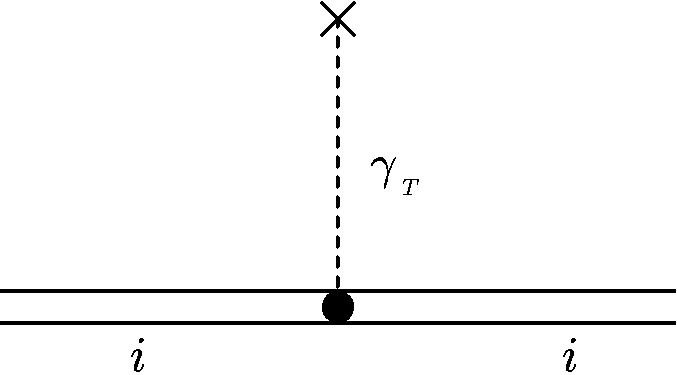}
\label{fig2}
\end{figure}

As a next step it is necessary to evaluate thermal radiative corrections of vertex type to the energy levels of final and initial states and their contribution to the transition rate. This correction corresponds to the Feynman graph depicted in Fig.~\ref{fig2}.

In \cite{S-2020} it was found that the thermal potential $V^\beta(r)$ can be derived within the framework of the rigorous QED theory. One of the advantages of the description in \cite{S-2020} is the ability to introduce thermal gauges in an obvious manner. Moreover, the regularization procedure proposed in \cite{S-2020}, which has eliminated divergences discussed above, led to obtaining the thermal Coulomb potential in closed form:
\begin{eqnarray}
\label{regular}
V^\beta(r)=\frac{4}{\pi}\left[-\frac{\gamma}{\beta}+\frac{i}{2r}\mathrm{ln}\frac{\Gamma\left(1+\frac{i r}{\beta}\right)}{\Gamma\left(1-\frac{i r}{\beta}\right)}\right],
\end{eqnarray}
where $\Gamma(z)$ is the gamma function. At room temperature and the low-lying states Eq. (\ref{regular}) could be approximated with sufficient accuracy by the regular term of Eq. (\ref{33}). 

Then, the thermal shift corresponding to the diagram in Fig.~\ref{fig2} for the hydrogen atom in the nonrelativistic limit and the point-nucleus approximation, can be found as 
\begin{eqnarray}
\label{shift}
\Delta E_{A}^{\beta} \equiv \langle A | V^\beta(r) | A\rangle  =
\\
\nonumber
=-\frac{4Ze^2\xi(3)}{3\pi\beta^3}\frac{n_{A}^2}{2}
[5n_{A}^2+1-3l_{A}(l_{A}+1)]a_{0}^2
,
\end{eqnarray} 
where $n_{A}$ is the principal quantum number of the hydrogenic state $A$, $l_{A}$ is the corresponding angular momentum, and $a_{0}$ is the Bohr radius.  It is should be noted that at room temperature the correction to the transition energies is on the level of precise measurement of $2s-1s$ transition frequency \cite{10Hz}. Recently lowest order thermal shift corresponding to thermal potential Eq. (\ref{33}) was also calculated for low-lying states in the helium atom, see \cite{SZA-2020}, where this correction was found on the level of the most precise measurements \cite{UltrahighPRL}.

This approximation, however, is violated for the Rydberg states. Numerical comparison of energy shifts $\Delta E^{\beta}_{A}=\langle A |V^\beta(r)| A\rangle$ for different atomic states $A$ calculated for both Eqs. (\ref{regular}) and (\ref{shift}) is given in Table~\ref{tab:comparison}.
\begin{center}
\begin{table}
\caption{Numerical values of energy shifts $\Delta E^{\beta}_{A}=\langle A |V^\beta(r)| A\rangle$ for different atomic states $A$ at temperatures $T=300$ K (upper line) and $T=3000$ K (lower line) in hydrogen atom. The first column shows the considered state $(n_{A},l_{A})$. In the second column the energy shift is calculated with approximate potential $V^\beta(r)$ given by Eqs. (\ref{33}) and (\ref{shift}). In the third column energy shift is calculated with potential $V^\beta(r)$ given by Eq. (\ref{regular}). All values are in Hz. }
\begin{tabular}{l c c}
\hline
\hline
$ (n_{A},l_{A}) $  & $\Delta E^{\beta}_{n_{A}l_{A}} $, Eq. (\ref{33})  & $ \Delta E^{\beta}_{n_{A}l_{A}}  $, Eq. (\ref{regular}) \\
\hline
(1,0)  & $-3.36$ & $-3.36$\\
       & $-3.36\times 10^{3}$&  $-3.35\times 10^{3}$\\
(2,0)  & $-46.98$ & $-46.98$ \\
       & $-4.7\times 10^4$ & $-4.68\times 10^4$\\
(10,0) & $-2.80\times 10^4 $ & $-2.76\times 10^4$\\
       & $-2.80\times 10^7$& $-1.35\times 10^7$\\      
(10,9) & $-1.29\times 10^4$ & $-1.28\times 10^4$\\
       & $-1.29\times 10^7$& $-8.50\times 10^6$\\   
(20,0) & $-4.48\times 10^5$ & $-3.68\times 10^5$\\
       & $-4.48\times 10^8 $& $-1.81\times 10^7$ \\ 
(20,19)& $-1.93\times 10^5$ & $-1.78 \times 10^5$\\
       & $-1.93\times 10^8$& $ -1.64\times 10^7$\\  
(100,0)& $-2.80\times 10^8$ & $ -1.79\times 10^{6}$\\
       & $-2.80\times 10^{11}$& $-1.79\times 10^7$\\   
(100,99)& $-1.14\times 10^8$& $-1.78 \times 10^6$\\
       & $ -1.14\times 10^{11} $& $ -1.79\times 10^7 $\\                          
\hline
\hline
\end{tabular}
\label{tab:comparison}
\end{table}
\end{center}

In particular, from Table~\ref{tab:comparison} it follows that Eq. (\ref{33}) is a good approximation for low-lying states at room temperatures, while for Rydberg states the accuracy and temperature behavior of Eq. (\ref{33}) fall out. Thus, the complete form, Eq. (\ref{regular}), should be used to evaluate the decay rates of highly excited states.

The energy shift defined by Eq. (\ref{shift}) results in a corresponding correction to the transition frequency and, consequently, to the transition rate. The latter can be written as the difference between Eq. (\ref{10a}) calculated with the zero-order transition frequency $\omega_{if}=E_{i}-E_{f}$ and Eq. (\ref{10a}) calculated with the corrected transition frequency $\widetilde{\omega}_{if}=E_{i}+\Delta E^{\beta}_{i} - E_{f}- \Delta E^{\beta}_{f}$, see \cite{ZSL-1ph, shabaev-report}:
\begin{eqnarray}
\label{29b}
\Delta W^{\mathrm{v}}_{if}=\frac{e^2}{2l_{i}+1}\sum_{m_{i}m_{f}}\frac{4}{3}(\omega^3_{if}-\widetilde{\omega}^3_{if})
|\langle f |\vec{r}|i\rangle|^2.\qquad
\end{eqnarray}
The calculated values of the correction Eq. (\ref{29b}) are collected in Table \ref{tab:1}. From Table \ref{tab:1} it is follows that corrections $\Delta W^{\mathrm{rad}}_{if}$ and $\Delta W^{\mathrm{v}}_{if}$ are of the same order.

In addition to the vertex correction to a spontaneous transition, $\Delta W^{\mathrm{v}}_{if}$, the corresponding correction to the induced transition rate, $\Delta W^{\mathrm{v,ind}}_{if}$, should be also evaluated. This correction can be obtained by multiplying Eq. (\ref{29b}) by the factor $n_{\beta}(\omega)$ taken at the appropriate frequency. The difference between induced decay rates, Eq. (\ref{10b}) with the energies of 'zero-order' and corrected is given by the expression
\begin{eqnarray}
\label{30b}
\Delta W^{\mathrm{v,ind}}_{if}=\frac{e^2}{2l_{i}+1}\sum_{m_{i}m_{f}}\frac{4}{3}|\langle f |\vec{r}|i\rangle|^2\times
\\
\nonumber
\left[\omega^3_{if}n_{\beta}(\omega_{if})-\widetilde{\omega}^3_{if}n_{\beta}(\widetilde{\omega}_{if})\right].
\end{eqnarray}

Numerical results for the spontaneous, induced transition rates and corrections $\Delta W^{\mathrm{v}}_{if}$ and $\Delta W^{\mathrm{v,ind}}_{if}$ are given in Table~\ref{tab:2d}. This combination makes it possible to visually assess the contribution of the vertex-type thermal corrections to the decay rates. It should be noted that the closed-form of the thermal potential Eq. (\ref{regular}) was used to estimate the corrections Eqs. (\ref{29b}) and (\ref{30b}) to the decay rates of highly excited states.
\begin{widetext}
\begin{center}
\begin{table}
\caption{Transition rates and thermal corrections at $ T=300 $ K to one-photon electric dipole transitions between highly excited states due to the thermal energy shift, see Eqs. (\ref{29b}), (\ref{30b}). All values are given in  s$^{-1}$.}
\begin{tabular}{l l c c c c}
\hline
\hline
$n_{i},l_{i}$ & $n_{f},l_{f}$ & $ W_{if} $ & $ \Delta W^{\rm ind}_{if} $ & $\Delta W^{{\rm v}}_{if}$ & $\Delta W^{{\rm v, ind}}_{if}$ \\
\hline
$(10,9)$ & $(9,8)$ & $ 1.320\times 10^{4} $ & $ 5.419\times 10^{3} $ & $ 2.190\times 10^{-5} $ &  $ 3.772\times 10^{-6} $\\

$ (50,1) $ & $(49,0)  $ & $ 2.682 $& $ 3.077\times 10^{2} $ & $ 6.780\times 10^{-6} $ & $ 5.173\times 10^{-4} $\\

$ (50,49) $ & $(49,48)  $ & $ 7.137\times 10^{-1} $& $81.861$ & $ 3.406\times 10^{-6} $ & $ 2.599\times 10^{-4} $\\

$ (70,1) $ & $(69,0)  $ & $ 4.840\times 10^{-1} $& $1.541\times10^{2}$ & $ -5.779\times 10^{-6} $ & $ -1.226\times 10^{-3}$\\

$ (70,69) $ & $(69,68)  $ & $ 9.369\times 10^{-2} $& $29.830$ & $ 1.633\times 10^{-6} $ & $ 3.464\times 10^{-4}$\\

$ (100,1) $ & $(99,0)  $ & $ 7.953\times 10^{-2} $& $ 74.387$ & $ 3.917\times 10^{-4} $ & $ 2.444\times 10^{-1} $\\

$ (100,99) $ & $(99,98)  $ & $ 1.093\times 10^{-2} $& $ 10.221$ & $ -1.972\times 10^{-7} $ & $ -1.229\times 10^{-4} $\\

\hline

\hline
\hline
\end{tabular}
\label{tab:2d}
\end{table}
\end{center}
\end{widetext}

\section{Discussion and conclusions}
\label{theend}

In the recent decades, photon emission processes have become of high interest in fundamental investigations on field theories, astrophysics, laboratory experiments, constructing of atomic clocks \cite{RubChSun,Eng,Feng,Martin,ZSL-1ph,jphysb2017}. The one-photon transitions play a special role in experiments pursuing the goal of precision determination of the fundamental physical constants \cite{H-exp}, and forbidden (magnetic dipole) one-photon transitions have found their application in atomic clocks, see, for example, \cite{AtCl-Cs}. To increase accuracy in all such experiments, the influence of the thermal environment should be taken into account. 

The most known phenomenon that affects the transition rate, is the blackbody radiation induced decays \cite{GC, Farley}. Basically, BBR-induced transitions are calculated in the framework of the quantum mechanical approach, while, as was recently shown in \cite{SLP-QED}, the application of the QED theory is more appropriate for detecting obscure effects arising in emission processes (an accurate accounting for the finite lifetimes of the excited states, for example). The QED approach allows the revealing the thermal effects which correspond to the known Feynman graphs with the replacement of the ordinary photon propagator by the thermal one \cite{S-2020}. For example, it has recently been shown that thermal self-energy radiative corrections to spontaneous one- and two-photon transition rates in the hydrogen atom, evaluated within the framework of rigorous QED theory at finite temperature \cite{ZSL-1ph, twophotonT}, are of particular importance in this field. 

In particular, it was demonstrated in \cite{ZSL-1ph} that the thermal radiative corrections to the spontaneous Ly$_\alpha$ decay rate can dominate over an ordinary induced transition rate up to temperatures $T<6000$ K. As a result of calculations in \cite{ZSL-1ph}, a contribution $ 2.4210\times 10^{-3} $ s$^{-1}$ at room temperature was found. From Table~\ref{tab:1} it follows that the total vertex-type thermal correction to the Ly$_\alpha$ decay rate is $ 3.748\times 10^{-5}$. At other temperatures the total thermal correction for low-lying states could be easily obtained by the multiplying by the factor $(T/300)^3$. As a consequence, the vertex-type thermal correction reaches the value $3.748\times 10^{-2}$ s$^{-1}$ at $T=3000$ K and $1.388 $ s$^{-1}$ at $10^4$ K for the Ly$_{\alpha}$ transition in the hydrogen atom, while the thermal self-energy correction found in \cite{ZSL-1ph} is $2.911 \times 10^{-1}$ s$^{-1}$ and $3.35\times 10^{-1}$ s$^{-1}$, respectively. Thus, we can conclude that vertex-type thermal correction can also exceed the induced transition rates for certain transitions and, therefore, is important in the astrophysical context. 
 
Thermal corrections of the vertex type can also be compared with the two-photon E1E2 and E1M1 transitions \cite{LabShonSol}. For the hydrogen atom and the Ly$_{\alpha}$ transition probability, the total contribution of these corrections exceeds the aggregated contribution of the corresponding two-photon transitions. On the other hand, calculations for H-like ions show that with increasing nuclear charge $Z$, the vertex-type corrections become less important and even negligible for high $Z$, see Table~\ref{tab:2}. The same conclusion follows for the thermal corrections to the forbidden M1 transitions collected in Table~\ref{tab:3}. However, the most interesting results arise for transitions between highly excited (Rydberg) states.
 
The relatively long lifetimes of Rydberg states make them suitable candidates for the implementation of quantum computers \cite{PhysRevLett.85.2208,PhysRevLett.104.010502,PhysRevLett.104.010503}. Thus, studying the accompanying effects can be important for their development. In particular, from Table~\ref{tab:2d} it follows that the vertex-type correction has the greatest effect on highly excited states. This conclusion seems obvious in conjunction with the result Eq. (\ref{shift}), which shows an increase in thermal correction with the principal quantum number of the excited state. However, the results of numerical calculations of the thermal correction Eq. (\ref{shift}), listed in Table~\ref{tab:comparison}, demonstrate the need to use the potential as a whole (without the series expansion over small parameter $r/\beta$) for highly excited states. Moreover, the numerical calculations of contributions corresponding to Fig. \ref{fig1} are nontrivial for such states within the B-splines method used in this work. The problem arises due to the summation over the entire spectrum and extremely large basis set for the qualitative approximation of final, initial, and intermediate states. Over methods such as Coulomb Green functions will also face this problem. 

The total thermal corrections to the transition rate should also include radiative self-energy corrections considered in \cite{ZSL-1ph}, which have double summation and more complicated analytical form in respect to Eq. (\ref{relmain}). Thus, the calculations of thermal corrections to the decay rates of highly excited states represent a separate task requiring the application of special methodology. However, the vertex-type correction to the transition rate arising due to the thermal shift of energy levels has a simpler representation, see Eqs. (\ref{29b}), (\ref{30b}), in the nonrelativistic limit. The corresponding contribution is the same order as the correction to wave function given by Fig. \ref{fig2}. Therefore, the vertex-type thermal correction to the transition rates between Rydberg states can be roughly estimated via the corrections given by the expressions (\ref{29b}) and (\ref{30b}). The results of numerical calculations of Eqs. (\ref{29b}), (\ref{30b}) at room temperature are listed in Table~\ref{tab:2d}. From Table~\ref{tab:2d} it follows that thermal corrections Eq. (\ref{29b}) are at the level of a few percent to the spontaneous transition rate for states with high angular momenta in the hydrogen atom. Although, the probability of stimulated transitions remains dominant.

Recently in \cite{newboson} it was suggested that precise measurements of particular transitions and its comparison with theoretical calculations could give rise in a prediction of possible new bosons (axions) beyond the standard model. It is suggested that these particles interact with ordinary matter through some potential. Different models of this potential and corresponding corrections to the atomic energy levels were discussed in the literature during the last decade \cite{axionflambaum,axionflambaum2,axionflambaum3}. However, for the theoretical predictions, the proper account of thermal shifts to energy levels is needed. In \cite{newboson} the analysis of such shifts was restricted by the BBR-induced Stark effect. In the present paper, it is demonstrated that, along with the BBR-induced Stark contribution, the vertex-type correction to the energy levels and transition rates determined by the thermal potential $V^{\beta}(r)$ should also be taken into account for future constraints on 'new physics'.

\section{Acknowledgements}
This work was supported by Russian Science Foundation (Grant No. 17-12-01035).

\bibliography{mybibfile} 

\end{document}